\documentstyle[aps,twocolumn,psfig,epsf]{revtex}
\author{Annalisa Fierro$^a$,  Mario Nicodemi$^{a,b}$ and Antonio Coniglio$^a$}
\address{
\vspace{0.2cm}
$^a$ Dipartimento di Fisica, Universit\'{a} ``Federico II'',
INFM and INFN Napoli, Via Cintia, 80126 Napoli, Italy\\
$^b$ Department of Mathematics,
Imperial College, London, SW7 2BZ, U.K.
}

\title{Equilibrium Distribution of the Inherent States and \\
their Dynamics in Glassy Systems and Granular Media}
\newcommand{\lan}{\langle}
\newcommand{\ran}{\rangle}
\newcommand{\e}{\mbox{e}}

\begin{document}

\maketitle
\bigskip
\begin{abstract}
The present paper proposes a Statistical Mechanics approach to the inherent
states of glassy
systems and granular materials, following the original ideas developed
by Edwards for granular materials.
Two lattice models, a diluted Spin Glass and a system of hard-spheres under
gravity, introduced in the context of glassy systems and granular materials,
are evolved using a  ``tap dynamics'' analogous to that of experiments
on granular materials.
The asymptotic macrostates, reached by the system, are shown
to be described by a single thermodynamical parameter, and this parameter to
coincide with the temperature, called the ``configurational
temperature'', predicted assuming that the distribution among the inherent
states satisfies the principle of maximum entropy.



\end{abstract}

\bigskip

The thermodynamics of macroscopic systems evolving
at equilibrium is well described
by Statistical Mechanics. However there are many systems, typically found
in ``frozen states'', where they do not evolve at all. These
are, for example, supercooled liquids quenched at zero temperature
in states, called inherent states \cite{Stillinger,sciortino},
corresponding to the local  minima of the potential energy in the
3N-dimensional configuration space of particle coordinates.
Granular materials \cite{JNBHM} at rest are another important example
of system frozen in mechanically stable microstates.
Grains are ``frozen'' because, due to their large masses \cite{JNBHM}, the
thermal kinetic energy is negligible compared to the gravitational energy;
thus the external bath temperature, $T_{bath}$, can be considered equal to
zero (by analogy with supercooled liquids, we call these mechanically
stable configurations inherent states).

In this paper, following the original ideas by Edwards for granular
materials \cite{Edwards} we attempt to develop a unified Statistical
Mechanics approach for the inherent states of glassy systems and granular
materials along the line of Ref. \cite{NC}.
The connection between Edwards approach and recent developments on glass
theory has received much attention
\cite{barrat,brey,Dean,berg,cugliandolo,Makse,Colizza}.
                       
The first step is to introduce a suitable dynamics which allows to explore
the configurations of the inherent states.
In granular materials the dynamics, from one stable microstate to another,
can be  induced by sequences of ``taps'', in which the energy is pumped
into the system in pulses. Due to inelastic collisions the kinetic energy
is totally dissipated after each tap, and the system is again frozen in
one of its inherent states \cite{Knight}. Similarly, in glass formers at
zero temperature the dynamics, from one inherent state to another, can be
induced by sequences of taps, where each tap consists in raising the bath
temperature and, after a lapse of time $\tau_0$, quenching it back to zero.
By repeating the process cyclically the system explores the space of the
inherent states \cite{NC,brey,Dean,berg,mehta,NCH}.
For a tap of infinite length ($\tau_0\rightarrow\infty$) the way to
explore the inherent states coincides with the one used in
\cite{Stillinger,sciortino} for a system of Lennard Jones mixture.
In the approach of Barrat {\em et al.} 
\cite{barrat} the system instead evolves in an out
of equilibrium quasi-stationary state at an external very low bath
temperature. In the limit of zero external temperature the system explores
the inherent states. 

Here we consider  a diluted Spin Glass and a system of hard-spheres under
gravity, introduced in the context of glassy systems and granular
materials, which are  evolved using a ``tap dynamics''. We show  that the
systems reach a stationary or quasi-stationary state which can be
characterized by a single thermodynamical parameter, defined through the
static fluctuation-dissipation relation, and this parameter coincides with
the ``configurational temperature'', predicted by the Edwards' hypothesis
of a flat measure for the microstate distribution. We also show that time
averages over the dynamics can be replaced by ensemble averages over such
measure.

\bigskip

\noindent

We first consider the Frustrated Lattice Gas model in three dimensions ($3D$).
The model was recently introduced to describe glass formers \cite{dfnc} and, 
in presence of gravity, granular materials \cite{NCH,our_rev,jef}.
It is made of particles with a twofold orientation (i.e., pointing in only two 
allowed directions), displaced on a cubic lattice (of linear size $L=8$ and  
overall density $\rho=\sum_i n_i/L^3=0.65$ and $0.75$),
and interacting via a quenched potential:
\begin{equation}
{\cal H}=J\sum_{\langle ij\rangle } f_{ij}(S_i,S_j)n_i n_j,
\label{FLG}
\label{H}
\end{equation}
where $n_i=0,1$ whether site $i$ is empty or filled by a particle,
$S_i={\pm} 1$ is a variable associated to the particle orientation,
$J$ the amplitude of the interaction potential between neighbours.
The shape factor $f_{ij}(S_i,S_j)=1-\epsilon_{ij}S_iS_j$ (where
$\epsilon_{ij}={\pm} 1$ are quenched and random variables) is 0 or 1
depending whether
the relative orientations $(S_i,S_j)$ are favoured or not when neighbouring
sites $i$ and $j$ are both occupied (for a review see \cite{our_rev}).

\begin{figure}[ht]
\centerline{
\hspace{-2.5cm}
\psfig{figure=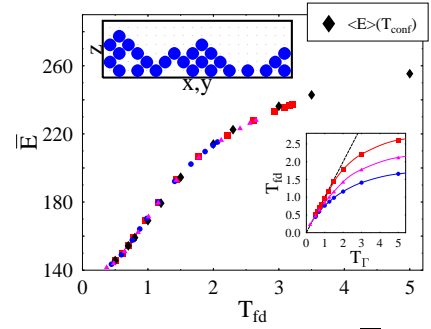,height=6.5cm,angle=-90}}
\vspace{-1.5cm}
\caption{
{\bf Main frame} The time average, ${\overline E}$, and the
ensemble average over the distribution eq.(\ref{pr}),
$\lan E \ran$ (black filled diamonds), plotted respectively
as a function of $T_{fd}$  and $T_{conf}$,
in the 3D hard-sphere system under gravity described in the text
(and schematically depicted in the {\bf upper inset}).
{\bf Lower Inset}
The temperature $T_{fd}\equiv \beta^{-1}_{fd}$ defined by
eq.(\ref{Sb}) as function of $T_\Gamma$
for $\tau_0=500,10,5~MCS$ (from top to bottom).
The straight line is the function $T_{fd}=T_\Gamma$.
}
\label{hard_sph}
\end{figure}

We also consider a 3D system of hard-spheres subject to gravity, where the
centers of mass of grains are constrained to move on the sites of a cubic
lattice, as depicted in the upper
inset of Fig.\ref{hard_sph}. Its Hamiltonian is
given by eq.(\ref{FLG}), plus a gravitational term, with $f_{ij}=1$ if $i$
and $j$ are nearest neighbours, $f_{ij}=0$ elsewhere and
$J\rightarrow\infty$ (in this case grains have no orientation
and $S_i$ is redundant $\forall i$).

In both models
the value of particle density is fixed, and a Monte Carlo tap
dynamics, which allows the system
to explore its inherent states, is applied.
During the dynamics,
the system cyclically evolves for a time $\tau_0$
(the tap duration \cite{nota1}) at a finite value of the bath temperature,
$T_{\Gamma}$ (the tap amplitude), and it is suddenly frozen at zero
temperature in one of its
inherent states (at zero temperature the system does not evolve anymore if the
energy cannot be decreased by one single particle movement).
After each tap, when the system is at rest, we record the
quantities of interest. The time, $t$, considered is therefore discrete and
coincides with the number of taps.

\bigskip

Let's first discuss the results about the Frustrated Lattice Gas model 
for density, $\rho=0.65$, since very similar features are
found for $\rho=0.75$ and in the hard-sphere system (described later on).

\begin{figure}[ht]
\centerline{\hspace{-2.5cm}
\psfig{figure=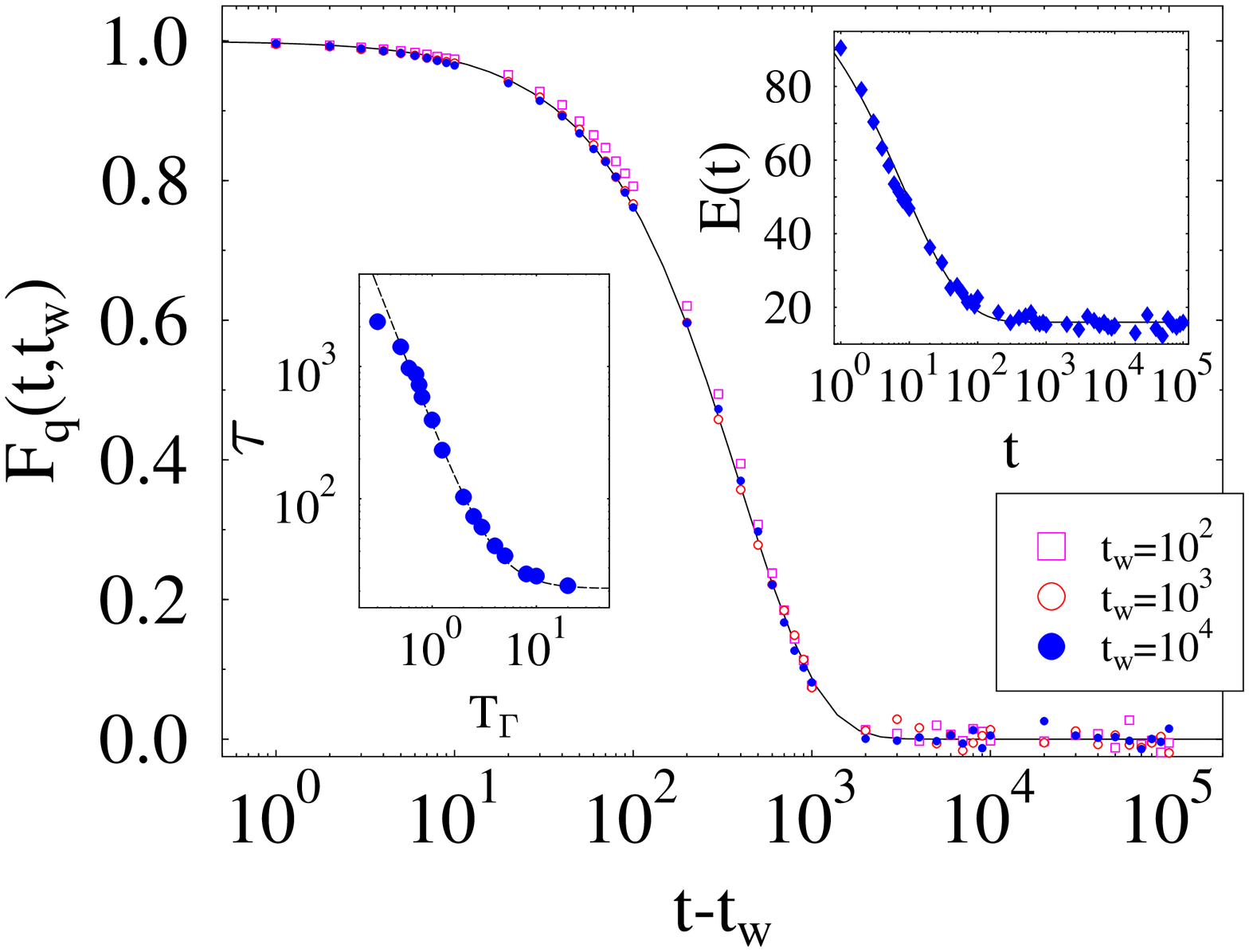,height=6.5cm,angle=-90}}
\vspace{-1.3cm}
\caption{{\bf Upper inset} The energy, $E(t)$,
of inherent states as a function of the ``tap'' number $t$, during a
tap dynamics with a tap amplitude $T_{\Gamma}=1.25~J$ and tap duration
$\tau_0=1~MCS$, in the Frustrated Lattice Gas of the text.
{\bf Main frame} The self-scattering two times
function, $F_q(t,t_w)=\sum_i\e^{\vec q{\cdot}[\vec r_i(t)-\vec r_i(t_w)]}/\rho
L^3$ as a function of the number  $t-t_w$
($T_{\Gamma}=1~J$, $\tau_0=1~MCS$ and $q=\pi/4$).
{\bf Lower Inset} The equilibration time, $\tau$, as a function of
$T_{\Gamma}$ (for $\tau_0=1~MCS$): $\tau$ diverges at low $T_{\Gamma}$.
After a transient,
$E(t)$ reaches its time independent asymptotic value, and $F_q=F_q(t-t_w)$
depends only on the difference of $t-t_w$, showing that the system has reached
a stationary state (our data are averaged up to $32$ noise realizations).}
\label{dinamica}
\end{figure}

\begin{figure}[ht]
\centerline{
\hspace{-2.5cm}
\psfig{figure=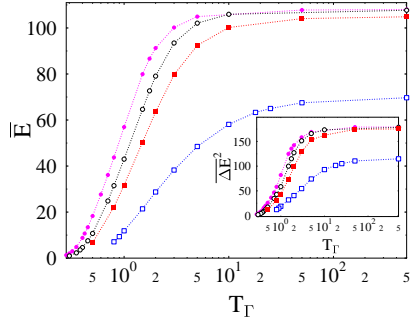,height=6.5cm,angle=-90}}
\vspace{-1.3cm}
\caption{The time average of the energy, ${\overline E}$, and ({\bf inset}) its
fluctuations, $\overline{\Delta E^2}$, recorded at stationarity during
a tap dynamics
with tap amplitude, $T_\Gamma$, in the Frustrated Lattice Gas model.
The four different curves correspond to sequences of tap with
different values of the duration of each single tap,
$\tau_0=100,~10,~5,~1~MCS$ (from top to bottom).
}
\label{energy}
\end{figure}

\begin{figure}[ht]
\centerline{
\hspace{-2.5cm}
\psfig{figure=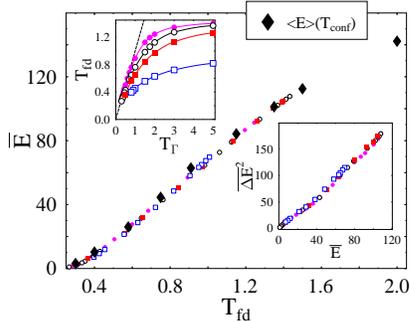,height=6.5cm,angle=-90}}
\vspace{-1.2cm}
\caption{{\bf Lower Inset} The time averages of energy fluctuations,
$\overline{\Delta E^2}$, when plotted as function of the time average of
energy, ${\overline E}$,
collapse on a single master function for all the different
values of tap amplitude and duration, $T_\Gamma$ and $\tau_0$, shown
in Fig.\ref{energy}.
{\bf Main frame} The time average, ${\overline E}$, and the
ensemble average over the distribution eq.(\ref{pr}),
$\lan E \ran$ (black filled diamonds), plotted respectively
as a function of $T_{fd}\equiv \beta^{-1}_{fd}$  and $T_{conf}$, in the 
Frustrated Lattice Gas model.
{\bf Upper Inset}
The temperature $T_{fd}\equiv \beta^{-1}_{fd}$ defined by
eq.(\ref{Sb}) as function of $T_\Gamma$
for $\tau_0=100,~10,~5,~1~MCS$ (from top to bottom).
The straight line is the function $T_{fd}=T_\Gamma$.
}
\label{universal}
\end{figure}

Interestingly, under the tap dynamics the system reaches a stationary state 
(see Fig.\ref{dinamica}) for  all the values of $T_{\Gamma}$ (and $\tau_0$) 
we considered.                                      
During the tap dynamics, in the stationary state, we have calculated the
time average of the energy,
${\overline E}$, and its fluctuations, ${\overline
{\Delta E^2}}$. We show the results in Fig.\ref{energy} as
function the tap amplitude, $T_\Gamma$ (for several values of the tap
duration, $\tau_0$). Apparently, $T_{\Gamma}$ is not the right
thermodynamical parameter, since sequences of ``taps'' with different
$\tau_0$ give different values of ${\overline E}(T_{\Gamma},\tau_0)$ and
${\overline {\Delta E^2}} (T_{\Gamma},\tau_0)$.
However, if the stationary states corresponding to
different tap dynamics (i.e., different $T_{\Gamma}$ and $\tau_0$), are
indeed
characterised by a {\em single} thermodynamical parameter,
all the curves should collapse onto a single master function when
${\overline {\Delta E^2}}$ is parametrically plotted as function of
${\overline E}$.
This data collapse  is in fact found and shown in the lower inset of
Fig.\ref{universal}. This is a prediction which
could be easily checked in real granular materials (where one could consider
the density which is easier to measure than the energy).

In the Frustrated Lattice Gas for density, $\rho= 0.75$, as much as in the 
hard-sphere model, for low values of the tap amplitude,
$T_{\Gamma}$, the system reaches a quasi-stationary state, where
one time quantities decay as the logarithm of time.
In this case the average over the time is performed over a time
interval of the tap dynamics such that the energy is practically constant.
By performing then the same procedure described in the stationary case,
we find (see Ref. \cite{FNCL})
again a collapse of data as in Fig.\ref{universal}.                             

The thermodynamical parameter, $\beta_{fd}$, is defined apart from an
integration constant, $\beta_0$, through the fluctuation-dissipation
relation:
\begin{equation}
-\frac{\partial {\overline E} }{\partial \beta_{fd}}  =
{\overline {\Delta E^2}}.
\label{Sb}
\end{equation}
By integrating eq. (\ref{Sb}), $\beta_{fd}-\beta_0$ can be expressed as
function of ${\overline E}$, where the integration
constant $\beta_0$ can be determined
independently \cite{FNCL}. In Fig.\ref{universal}, ${\overline E}$ as
function of $T_{fd}\equiv\beta_{fd}^{-1}$ is shown. 
The corresponding results for
the hard-sphere system are shown in Fig.\ref{hard_sph}. In the last model
we have also checked that the system density on the bottom layer, $\rho_b$,
and the density self-overlap function, $Q$ 
depend only on $\beta_{fd}$ (see Ref. \cite{FNCL}),
confirming that a
unique thermodynamical parameter is enough to describe the system
macrostates.


We have found that the fluctuations of the energy
in the stationary state depend only on the energy, $\overline{E}$,
and not on the past
history. If all macroscopic quantities depend only on the energy,
$\overline{E}$,
or on its conjugate thermodynamical parameter, $\beta_{fd}$,
the stationary state
can be  genuinely considered a ``thermodynamical state''. If this is the case
one can attempt to construct an equilibrium statistical mechanics, as originally
suggested by Edwards \cite{Edwards}.

More precisely we ask in the stationary regime
what is the probability distribution, $P_r$,
of finding the system in the inherent state $r$ of energy $E_r$ (see
\cite{NC}).  We assume that the distribution is given by the principle of
maximum entropy, $S=-\sum_r P_r
ln P_r$, under the condition that the average energy is fixed: $E =
\sum_r P_r E_r $. Thus, we have to maximise the following
functional: $ I[P_r] =-\sum_r P_r
ln P_r -\beta_{conf} (E - \sum_r P_r E_r) $.  Here $\beta_{conf}$ is a
Lagrange multiplier determined by the constraint on the energy and takes
the name of ``inverse configurational temperature''. This procedure leads
to the Gibbs result:
\begin{eqnarray}
P_r=\frac{e^{-\beta_{conf} E_r}}{Z}
\label{pr}
\end{eqnarray}
where $Z=\sum_r e^{-\beta_{conf} E_r}$. Using standard Statistical Mechanics
it is easy to show that, in the
thermodynamic limit, the entropy $S$  and $\beta_{conf}$ are also given by:
\begin{eqnarray}
S = ln \Omega (E),\quad \beta_{conf}= \frac{\partial ln \Omega}{\partial E}
\label{omega}
\end{eqnarray}
where $\Omega (E)$ is the number of inherent states corresponding to
energy $E$.

It is possible to show \cite{NC} that in the particular case in which the
particles density $\rho$ is constant, $T_{conf}$ is simply related to the
``compactivity'', $X=\left(\frac{\partial S}{\partial V}\right)^{-1}$,
introduced by Edwards in his seminal papers \cite{Edwards}.
In general $X$ and $T_{conf}$ are two independent variables.

If the distribution in the stationary state coincides with eq.(\ref{pr})
the time average of the energy, ${\overline  E}(\beta_{fd})$, recorded
during the tap dynamics, must coincide with the ensemble average, $\lan
E\ran(\beta_{conf})$, over the distribution eq.(\ref{pr}). To calculate
the average $\lan E\ran$, as function of $\beta_{conf}$, we have
introduced an auxiliary hamiltonian (see also \cite{barrat}) ${\cal
H'}(\{S_i, n_i\}) = {\cal H}(\{S_i, n_i\}) +
\delta(\{S_i, n_i\})$, where  ${\cal H}(\{S_i, n_i\})$ is the
hamiltonian (\ref{FLG}), and $\delta(\{S_i, n_i\})$, is zero, if the
configuration is an inherent one, and infinite, otherwise. In this way the
canonical distribution for this Hamiltonian gives a weight,
$e^{{-\beta_{conf}}{\cal H'}}$, which is 
equal to $e^{{-\beta_{conf}}{\cal H}}$, for
the inherent configurations, and zero otherwise, reproducing the
distribution eq.(\ref{pr}). With this auxiliary hamiltonian, using
standard Monte Carlo simulations, we have then calculated $\lan
E\ran(\beta_{conf})$. Fig.s \ref{universal}, \ref{hard_sph} outline a very
good agreement between $\lan E
\ran(\beta_{conf})$ and ${\overline E}(\beta_{fd})$ (notice that there are
no adjustable parameters). The same agreement is found in the
Frustrated Lattice Gas model for $\rho=0.75$ and in the hard-sphere 
system for the other
quoted observables, $\rho_b$, and $Q$ (see Ref. \cite{FNCL}).

In the approach of Ref. \cite{barrat} the system
explores the inherent states evolving in an out of equilibrium quasi-stationary 
state at a very low bath temperature;
in this approach  the configurational temperature is 
expected  
to coincide with  the ``dynamical temperature'', $T_{dyn}$,
which appears in the extension of the fluctuation-dissipation relation to
the out-equilibrium case. One of the differences with the approach used 
here is that using the tap dynamics it is also possible to
explore low density inherent states in a stationary regime and not only
the off-equilibrium ``glassy regime''.   
For istance, the Frustrated Lattice Gas model at density $\rho=0.65$ 
(one of the cases here studied) is never found in an out
of equilibrium quasi-stationary state (at any finite value of the bath
temperature the system quickly reaches the equilibrium state).


In conclusion, in the context of models for glasses and granular
materials, we have obtained two different results. First, we have shown
that the stationary states reached by the system under the tap dynamics
among the inherent states are not dependent on the past history and can be
considered as a thermodynamical state characterized by a single
thermodynamical parameter, $T_{fd}$, defined through the fluctuation-dissipation
relation. Second,  $T_{fd}$ coincides with $T_{conf}$,
predicted assuming that the distribution among the inherent states
satisfies the principle of maximum entropy under the constraint that
energy is fixed. Moreover ensemble average coincides with time average
over the tap dynamics. In particular we have found that, by using $T_{conf}$
as a state parameter, the observables recorded in different tap sequences
(different amplitude and duration of taps) fall onto universal master
curves, and the curves turn out to coincide with the ones predicted by the
distribution eq.(\ref{pr}).


\bigskip

We thank S.F. Edwards for useful discussions.
This work was partially supported by the TMR-ERBFMRXCT980183,
INFM-PRA(HOP), MURST-PRIN 2000. The allocation of computer resources from INFM
Progetto Calcolo Parallelo is acknowledged.

\vspace{2cm}
\vspace{-1.5cm}

\begin{thebibliography}{40}
\bibitem{nota1} $\tau_0$ is measured in Monte Carlo steps (MCS), where $1~MCS$
corresponds to $N$ attempts to move a particle randomly chosen, and $N$ is the
number of particles.
\end{thebibliography}
\end{document}